\documentclass[twoside,twocolumn]{article}
\usepackage[super,sort&compress,comma]{natbib} 
\usepackage[version=3]{mhchem}
\usepackage{txfonts}
\usepackage{sectsty}
\usepackage[title, titletoc]{appendix}
\usepackage{balance} 
\usepackage{graphicx} 
\usepackage{graphbox}
\usepackage{lastpage}
\usepackage[format=plain,justification=raggedright,singlelinecheck=false,font=small,labelfont=bf,labelsep=space]{caption} 
\usepackage{fancyhdr}
\usepackage{bm}
\usepackage{float}
\usepackage{color}
\usepackage{array}

\begin{document}

\thispagestyle{plain}
\fancypagestyle{plain}{
\renewcommand{\headrulewidth}{1pt}}
\renewcommand{\thefootnote}{\fnsymbol{footnote}}
\renewcommand\footnoterule{\vspace*{1pt}%
\hrule width 3.4in height 0.4pt \vspace*{5pt}} 
\setcounter{secnumdepth}{5}

\makeatletter 
\def\subsubsection{\@startsection{subsubsection}{3}{10pt}{-1.25ex plus -1ex minus -.1ex}{0ex plus 0ex}{\normalsize\bf}} 
\def\paragraph{\@startsection{paragraph}{4}{10pt}{-1.25ex plus -1ex minus -.1ex}{0ex plus 0ex}{\normalsize\textit}} 
\renewcommand\@biblabel[1]{#1}            
\renewcommand\@makefntext[1]%
{\noindent\makebox[0pt][r]{\@thefnmark\,}#1}
\makeatother 
\renewcommand{\figurename}{\small{Fig.}~}
\sectionfont{\large}
\subsectionfont{\normalsize} 

\fancyfoot{}
\fancyfoot[RO]{\footnotesize{\sffamily{1--\pageref{LastPage} ~\textbar  \hspace{2pt}\thepage}}}
\fancyfoot[LE]{\footnotesize{\sffamily{\thepage~\textbar\hspace{3.45cm} 1--\pageref{LastPage}}}}
\fancyhead{}
\renewcommand{\headrulewidth}{1pt} 
\renewcommand{\footrulewidth}{1pt}
\setlength{\arrayrulewidth}{1pt}
\setlength{\columnsep}{6.5mm}
\setlength\bibsep{1pt}

\twocolumn[
  \begin{@twocolumnfalse}
\noindent\LARGE{\textbf{Anchoring-driven spontaneous rotations in active gel droplets}}
\vspace{0.6cm}

\noindent\large{\textbf{A.~R. Fialho\textit{$^{a}$}, M.~L. Blow{$^{a}$}, D. Marenduzzo\textit{$^{a}$}}}
\vspace{0.5cm}

\noindent\textit{\small{\textbf{Received Xth XXXXXXXXXX 20XX, Accepted Xth XXXXXXXXX 20XX\newline
First published on the web Xth XXXXXXXXXX 200X}}}

\noindent \textbf{\small{DOI: 10.1039/b000000x}}
\vspace{0.6cm}

\noindent \normalsize{We study the dynamics of an active gel droplet with imposed orientational anchoring (normal or planar) at its surface. We find that if the activity is large enough droplets subject to strong anchoring spontaneously start to rotate, with the sense of rotation randomly selected by fluctuations. Contractile droplets rotate only for planar anchoring and extensile ones only for normal anchoring. This is because such a combination leads to a pair of stable elastic deformations which creates an active torque to power the rotation. Interestingly, under these conditions there is a conflict between the anchoring promoted thermodynamically and that favoured by activity. By tuning activity and anchoring strength, we find a wealth of qualitatively different droplet morphologies and spatiotemporal patterns, encompassing steady rotations, oscillations, and more irregular trajectories. The spontaneous rotations we observe are fundamentally different from previously reported instances of rotating defects in active fluids as they require the presence of strong enough anchoring and entail significant droplet shape deformations.}
\vspace{0.5cm}
\end{@twocolumnfalse}
  ]

\footnotetext{\textit{$^a$School of Physics and Astronomy, University of Edinburgh, Edinburgh EH9 3FD, UK\\}}

\section{Introduction}

Active matter is an area of condensed matter physics which has witnessed tremendous development and an increase in attention over the past decade. Within this context, an ``active system'' is generally defined as an ensemble of ``active particles'' which continuously exert non-thermal (i.e., active) forces on the environment, for instance (but not necessarily) to move. A powerful framework within which to understand the physical properties of soft living and active matter has been provided by the hydrodynamic theory of active gels~\cite{kruse,cristinareview,sriramreview2}.

This theory starts from the acknowledgement that all active forces are internally generated -- there is no external force exerted on any particle -- hence they should sum up to zero. Thus, to lowest order, it is possible to treat each particle as an active force dipole. As a consequence, an active particle is naturally associated with a direction, that of its active dipole. The underlying physics and hydrodynamics are therefore those of liquid crystals~\cite{kruse}. Furthermore, in several cases the microscopic constituents, or particles, are elongated, hence orientational order arises even in the passive limit due to steric interactions. Indeed, instances of active gels include suspensions of bacteria, which are rod-like, or mixtures of molecular motors and actin or microtubules, which are stiff polymers. An important distinction is whether the forces in the active dipoles are directed from the centre of mass of the active particles outwards, or inwards. In the former case, the resulting active materials are named ``extensile'', in the latter ``contractile''~\cite{cristinareview,sriramreview2}. Examples of extensile active gels are the hierarchical active matter made up by kinesin walking on microtubule bundles of Ref.~\cite{sanchez}, or concentrated bacterial suspensions as in~\cite{goldstein}; examples of contractile active gels are actin-myosin suspensions as in~\cite{mackintosh}, or suspensions of algae such as {\it Chlamydomonas}. 

Here we are concerned with a binary system where a droplet of active gel is embedded in an isotropic and Newtonian, aqueous solvent. This geometry, previously considered in~\cite{elsen,giomi,elsen2}, can be realised in the lab through actomyosin droplets or cell extracts, and it could also be made by encapsulating kinesin-microtubule gels into oil-water emulsions, as done in~\cite{sanchez}. 

Previous work has shown that, in the absence of any imposed anchoring, such droplets can become self-motile if the magnitude of the active stress is large enough. The onset of motility is linked to the set-up of force multipoles~\cite{rhoda1,rhoda2}, as in other examples of active droplets which are not based on active gels~\cite{yoshinaga1,yoshinaga2,tutupalli,stark}.
The key control parameter in the simulations of~\cite{elsen} was the dimensionless activity parameter $\Theta=\zeta R^2/ \kappa$, where $\zeta$ is the active stress, $R$ is the droplet radius, and $\kappa$ is the elastic constant of the active gel. If $\Theta$ is small, the droplet deforms due to activity but it is stationary; instead, if $\Theta$ exceeds a critical threshold, then elastic deformations (splay or bend) create an imbalance in internal active forces which leads to motion. The transition between quiescent and self-motile phases requires the spontaneous breaking of the polarity inversion symmetry, and provides a mechanism through which cell extracts (viewed as actomyosin droplets with a surface tension) could, in principle, move solely by virtue of myosin contractility, and in the absence of actin treadmilling, which instead regulates standard cell crawling on a substrate~\cite{elsen2}. The underlying mechanism is essentially the same as that of the so-called ``generic instability'' of bulk active fluids~\cite{sriram,spontaneousflow,spontaneousflowLB,cristinareview}. Further work in~\cite{giomi} has shown that a droplet of contractile active nematic can even split spontaneously at large values of the activity. This work proposed that a second important dimensionless number should be an ``active capillary number'', measuring the ratio between active forces and surface tension $\tilde{\sigma}$, equal to $\zeta R/\tilde{\sigma}$. 

Here, instead, we focus on the dynamics of an active droplet in the presence of an imposed (e.g., thermodynamic) anchoring, which favours either normal or planar alignment of the polarisation field (i.e., the coarse grained average direction of the active dipoles) at the droplet surface. If the strength of the anchoring is large enough, we find that an increase in activity leads to spontaneously rotating, rather than self-motile, droplets. Spontaneous rotations are powered by elastic deformations and are typically accompanied by a change in droplet shape (droplets often attain a bean-shaped morphology). The nature of the transition between the quiescent and rotating phases depends on whether the droplet is contractile or extensile: hysteresis is found in the former case, but not in the latter one. Apart from a large enough dimensionless activity $\Theta$ (and possibly capillary number), we find that spontaneous rotations of the kind we observe require a sufficiently strong anchoring.

Our simulations further suggest that rotations require the anchoring to be of a particular nature: normal for extensile droplets and planar for contractile ones. A sufficiently strong anchoring of this kind is necessary to stabilise a pair of elastic deformations within the droplet: this polarisation pattern is associated with a pair of non-collinear active forces, which then creates a non-zero torque. 

Alignment at the surface is also affected by activity, as observed in previous simulations, in the absence of any thermodynamic anchoring. These simulations have shown that active flows generated at the interface can lead to a dominant type of interfacial alignment~\cite{Matthew} (the simulations were of nematic-isotropic interfaces, but the same effect occurs with polar order). Disregarding flow alignment effects, the interfacial alignment is planar for extensile activity and normal for contractile activity. This so-called ``active anchoring'' is produced by the combination of active normal stresses which deform the droplet, and active shear flows which turn the active dipole direction (see~\cite{Matthew} for more details).
Since active anchoring results from active stresses that extend into the bulk of the active fluid, its effective anchoring strength is $\sim\zeta l$, where $l$ is the typical lengthscale of distortions in the nematic phase~\cite{GiulioMatthew}.

Quite remarkably, for the cases we present here, we only observe droplet rotations when the nature of the thermodynamically imposed anchoring (i.e., whether planar or normal to the surface) conflicts with the nature of the effective anchoring that arises due to activity.  

In terms of dimensionless numbers, we therefore speculate that an additional useful quantity should be the ratio between the thermodynamic strength $w_t$ (measured in N/m) and the strength of active anchoring $w_a\sim\zeta l$. If the active anchoring $w_a$ dominates, upon increasing $\zeta$ the droplet becomes self-motile and translates spontaneously. Instead, if the thermodynamic anchoring $w_t$ dominates over $w_a$, an increase of $\Theta$ leads to spontaneous rotations. [A more detailed discussion of this additional dimensionless number is given in the next Section.]

\section{Model and methods}

We model the active gel droplet in the hydrodynamic (continuum) limit, describing the active polar liquid crystal in terms of a set of coarse grained fields. Our key hydrodynamic variables are: the concentration of active particles $\phi \left( \textbf{r} , t \right)$, the polarisation field $\textbf{P} \left( \textbf{r} , t \right)$ of the active liquid crystal phase \footnote{The polarisation field is defined as the mesoscopic average over all particle orientations $\textbf{P} = \langle \textbf{p} \rangle$, where $\textbf{p}$ is a unit vector describing the orientation of each particle.}, and the average velocity field $\textbf{v} \left( \textbf{r} , t \right)$ of both the particles and the solvent.

In the limit of zero activity, the passive system tends to minimize the free energy functional 
\begin{eqnarray}\label{FreeEn}
F \left[ \phi , \textbf{P}\right] &=& \int d^3 r\, \bigg\{ \frac{a}{4 \phi_{cr}^2} \phi^2 \left( \phi -\phi_0 \right)^2 + \frac{k}{2} | \nabla \phi |^2 \\ \nonumber
&-& \frac{\alpha}{2} \frac{(\phi - \phi_{cr})}{\phi_{cr}} |\textbf{P}|^2 + \frac{\alpha}{4} |\textbf{P}|^4 + \frac{\kappa}{2} \left( \nabla \textbf{P} \right)^2 + W \left( \textbf{P} \cdot \nabla \phi \right)^2 \bigg\}.
\end{eqnarray}
The first two terms of the free energy stem from the binary fluid theory and allow for the formation of a droplet, which is stabilised by interfacial tension. The first term creates a double well potential with two minima states: a polar active phase inside the droplet, and a passive isotropic phase outside it. The interfacial tension in between these two phases is determined by the second term, proportional to $k$. The remaining terms in Eq.~(\ref{FreeEn}) account for orientational order and are taken from liquid crystal theory. The terms proportional to $|\textbf{P}|^2$ and $|\textbf{P}|^4$ constitute the Landau potential of the polar phase and describe a second phase transition from isotropic $|\textbf{P}|=0$ to polar $|\textbf{P}| \neq 0$ at critical concentration $\phi_{cr}$. $\alpha > 0$ is the phenomenological free energy amplitude and controls the transition. The term in $\left( \nabla \textbf{P} \right)^2$ penalises elastic distortions from the local polar alignment, while $\kappa$ is the effective elastic constant within the single elastic constant approximation . The final term represents the anchoring of $\textbf{P}$ to the droplet interface. $W$ controls the anchoring strength and the type of anchoring. Planar alignment is favoured for $W>0$, and normal anchoring corresponds to $W<0$. Most of our results correspond to the strong anchoring regime, although we varied the value of $W$ in selected cases. We should note here that both $W$ and $k$ in Eq.~(\ref{FreeEn}) have units of N, whereas the quantities they are linked to, the experimentally measured strength of anchoring, which we call $w$, and the surface tension, $\tilde{\sigma}$, have units of $N/m$.

Assuming local conservation of active material, the concentration $\phi$ follows the following Cahn-Hilliard-like equation:
\begin{equation}
\frac{\partial \phi}{\partial t} + \textbf{v} \cdot \nabla \phi = M \nabla^2  \left( \frac{\delta F}{\delta \phi} \right)
\label{evolution_phi}
\end{equation}
where $M \nabla^2 \left( \frac{\delta F}{\delta \phi} \right) $ represents the diffusive current, with $M$ the thermodynamic mobility of the active particles, and $\frac{\delta F}{\delta \phi}$ the chemical potential derived from the free energy, Eq.~(\ref{FreeEn}). 

The dynamics of the polarisation field $\textbf{P} \left( \textbf{r} , t \right)$ follows from polar liquid crystal theory and can be written as 
\begin{equation}
\frac{\partial \textbf{P}}{\partial t} + \left( \textbf{v} \cdot \nabla \right) \textbf{P} = - \frac{1}{\Gamma} \frac{\delta F}{\delta \textbf{P}} - \underline{\underline{\Omega}} \textbf{P} + \xi \underline{\underline{\nu}} \textbf{P} .
\label{evolution_P}
\end{equation}
Eq.~(\ref{evolution_P}) is a convection-relaxation equation where $\underline{\underline{\nu}}$ and $\underline{\underline{\Omega}}$ are, respectively, the symmetric and anti-symmetric parts of the velocity gradient tensor $\nabla \textbf{v}$, $\Gamma$ is the rotational viscosity parameter, and $\frac{\delta F}{\delta \textbf{P}}$ is the molecular field $\textbf{h}$ resultant from Eq.~(\ref{FreeEn}). The parameter $\xi$ is a shape factor related to the aspect ratio of the active particle: $\xi > 0$ for rod-like particles, and $\xi < 0$ for disk-like particles. It also determines the particle's behaviour under flow shear: shear-alignment corresponds to $ | \xi | > 1$, and shear-tumbling to $ | \xi | < 1 $.

Momentum balance in the system is enforced through the incompressible Navier-Stokes equation
\begin{equation}
\nabla \cdot \textbf{v} = 0 ,
\label{incompressibility}
\end{equation}
\begin{equation}
\rho \left( \frac{\partial}{\partial t} + \textbf{v} \cdot \nabla \right) \textbf{v}  = - \nabla p + \nabla \cdot \underline{\underline{\sigma}} - \gamma \textbf{v} .
\label{Navier-Stokes}
\end{equation}
In Eq.~(\ref{Navier-Stokes}), $\rho$ is the constant mass density of the fluid, $p$ is the isotropic pressure and $- \gamma \textbf{v}$ the frictional force. $\underline{\underline{\sigma}}$ is the total hydrodynamic stress, which includes viscous, elastic, interfacial, and ``active'' contributions, written respectively as:
\begin{equation}
\sigma^{viscous}_{\alpha \beta} = \eta \left( \partial_{\alpha} v_{\beta} + \partial_{\beta} v_{\alpha} \right) ,
\label{sigma_viscous}
\end{equation}
\begin{equation}
\sigma^{elastic}_{\alpha \beta} = \frac{1}{2} \left( P_{\alpha} h_{\beta} - P_{\beta} h_{\alpha} \right) - \frac{\xi}{2} \left( P_{\alpha} h_{\beta} + P_{\beta} h_{\alpha} \right) - \kappa \partial_{\alpha} P_{\gamma} \partial_{\beta} P_{\gamma} ,
\label{sigma_elastic}
\end{equation}
\begin{equation}
\sigma^{interface}_{\alpha \beta} = \left( f - \phi \frac{\delta F}{\delta \phi} \right) \delta_{\alpha \beta} - \frac{\partial f}{\partial \left( \partial_{\beta} \phi \right)}  \partial_{\alpha} \phi  ,
\label{sigma_interface}
\end{equation}
\begin{equation}
\sigma^{active}_{\alpha \beta} = - \zeta \phi P_{\alpha} P_{\beta} .
\label{sigma_active}
\end{equation}
In Eq.(\ref{sigma_viscous}), $\eta$ is the shear viscosity and $f$, in Eq.(\ref{sigma_interface}), is defined as the free energy density. Greek indices represent cartesian coordinates. The elastic (Ericksen) stress and the interfacial stress stem, respectively, from liquid crystal and binary fluid formalisms. 
The active stress (\ref{sigma_active}) is derived by summing over the contributions from each force dipole and coarse-graining~\cite{hatwalne}; the resultant parameter $\zeta$ is the active parameter and it is positive for extensile particles (pushers), and negative for contractile particles (pullers).  The magnitude of the activity parameter is  proportional to the strength of the force dipoles.

The active droplets are simulated in two dimensions on a square lattice. The droplet configuration is initialized as a circular domain of radius $R$. In the interior of the droplet, the concentration is $\phi = \phi_0$ and the vector polarisation $\textbf{P}$ is vertically aligned with unit magnitude $|\textbf{P}|=1$. Outside of the droplet, both $\phi$ and $\textbf{P}$ are zero. To obtain spiral rotating droplets, the polarisation is initialised with the shape of an aster inside of the circular domain.

The equations that describe the system are then solved using a hybrid lattice Boltzmann method~\cite{spontaneousflowLB}. This involves solving Navier-Stokes equation (\ref{Navier-Stokes}) by means of a lattice Boltzmann algorithm, while Eq.~(\ref{evolution_phi}) and Eq.~(\ref{evolution_P}) are solved using finite difference methods.

Unless explicitly stated, the parameters used for the simulations are: $R=20$, $\phi_0=2$, $\phi_{cr}=1$, $a=0.04$, $k=0.25$ for extensile droplets and $k=0.06$ for contractile droplets, $\alpha=0.1$, $\kappa=0.04$, $\xi=1.1$ for flow alignment and $\xi=0.5$ for flow tumbling, $\eta=1$, $W=0.1$ for planar anchoring and $W=-0.05$ for normal anchoring.


All the parameter values are quoted throughout in simulation units. A mapping to physical units can be obtained by choosing appropriate scales for length, time and force. For concreteness, we chose $L=1 \mu m$,  $\tau=10 ms$, and $F=100 nN$. These values correspond to actomyosin (a contractile active gel). The correspondence between simulation and physical units is listed in table \ref{parameters}.
\begin{table}
\begin{tabular}{p{2.7cm}|c|c}
Model variables and parameters (actomyosin) & Simulation units & Physical units  \\
\hline 
Effective shear viscosity, $\eta$ 			 & $1$  		 & $1\, \mathrm{kPa}\cdot \mathrm{s}$ \\
Effective elastic constant, $\kappa$ 			 & $0.02-0.08$ 		 & $2-8\, \mathrm{nN}$ \\
Shape factor, $\xi$ 					 & $1.1$, $0.5$		 & dimensionless \\
Effective diffusion constant, $D=Ma$			 & $0.004$ 		 & $0.4\,\mu \mathrm{m}^{2}/\mathrm{s}$ \\ 
Rotational viscosity, $\Gamma$ 				 & $1$ 			 & $1\, \mathrm{kPa}\cdot \mathrm{s}$ \\
Activity, $\zeta$ 					 & $0-0.01$ 		 & $0-1\, \mathrm{kPa}$ \\
Anchoring strength, $W$             & $0-0.1$            & $ 0-10\, \mathrm{nN}$ \\   
\end{tabular}
\caption{Model parameters in simulation units and the correspondent physical units (for an actomyosin gel:$L=1 \mu m$,  $\tau=10 ms$, and $F=100 nN$).}
\label{parameters}
\end{table}

It is also useful to list here the key dimensionless numbers in our model, which will be referred to when discussing the results. First, $\Theta=\frac{\zeta R^2}{\kappa}$ measures the relative importance of active and elastic stresses; in our case, it controls the onset of active spontaneous flow within the droplet. Second, there is an active capillary number, $\Phi=\frac{\zeta R}{\sigma}$, where $\sigma$ is the surface tension of the droplet: this captures the relative importance of active and interfacial forces, and, in our case, it controls the extent of droplet deformation. A third dimensionless number is the ratio between the strength of the thermodynamic and active anchoring. Dimensional analysis suggests that the strength of the thermodynamic anchoring, $w_t$, can be estimated as $W/l_i$, where $l_i$ is an interfacial lengthscale. If the bulk energy density scale $a$ and $\alpha$ are comparable (as in our simulations), we expect $l_i\sim \sqrt{k/a}$ for $W\ll K$, and $l_i\sim \sqrt{Wa}$ for $W\gg K$. On the other hand, as anticipated in the introduction, previous work has shown that the active anchoring can be estimated by $\sim \zeta l$, with $l$ the lengthscale of elastic distortions in the polarisation field.

\section{Results}

\subsection{Anchoring-driven spontaneous rotation of extensile droplets}

We begin by studying the dynamics of a droplet of extensile active gel, with initially uniform polarisation (see Fig.~1): we refer to this as the ``aligned initial condition''. [Recall that the extensile case is relevant in practice to the ``hierarchically assembled active matter'' of Ref.~\cite{sanchez} made up by microtubules and molecular motors.]

\begin{figure*}[htb]     
        \begin{center} 
        \includegraphics[width=\textwidth]{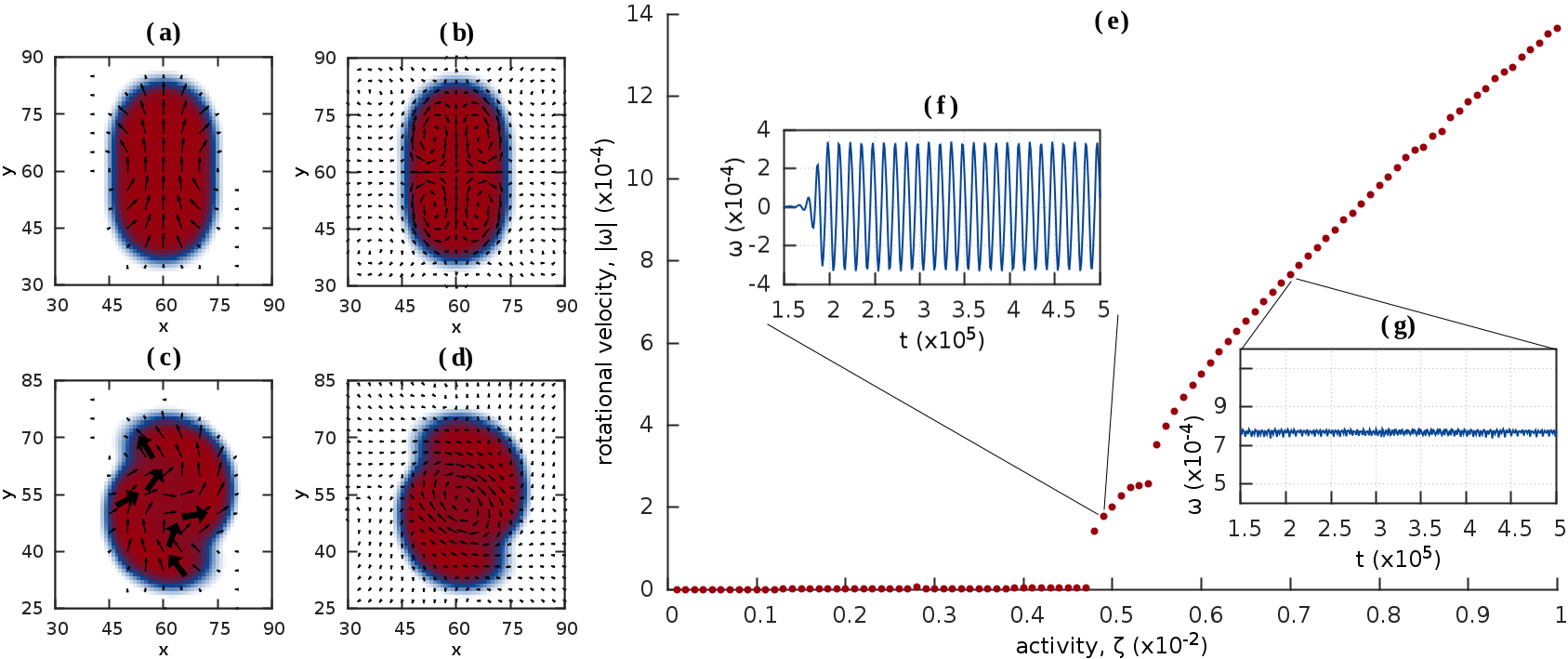}
\end{center}
\caption{Spontaneous rotation of extensile droplets with strong normal anchoring and aligned initial conditions. (a)-(d) Snapshots of the droplet shapes and polarisation field (a),(c), or velocity field (b),(d). The droplet in (a),(b) is quiescent, whereas in (c),(d) it rotates spontaneously in a clockwise sense. (e) Plot of the rotational velocity of the droplet as a function of activity. Insets in (f),(g) show the time dependence of the rotational velocity during a simulation.}
\label{fig1}                 
\end{figure*}

In Figure 1 we consider normal anchoring of the active dipole orientation at the droplet boundary (with $W=-0.05$, see previous Section) -- we briefly discuss the case of extensile gels with planar anchoring below. For small values of the activity, the droplet elongates slightly and becomes elliptical in steady state, with long axis along the direction of polarisation (Fig.~1(a)). While the droplet is static, there is an internal flow with the shape of four symmetric vortices (Fig.~1(b)): these flows are driven by the elastic deformation (mainly bend) that is induced by the anchoring.

For large enough activity, we find that the droplet substantially deforms (Fig.~1(c)) and can rotate spontaneously (Fig.~1(d), Suppl. Movie  1). In the spontaneously rotating phase, the internal flow field has the topology of a single vortex (Fig.~1(d)). The sense of rotation of the droplet can be understood by looking at the patterns of bend deformation in the polarisation (Fig.~1(d)): these distortions lead to active forces which push the droplet in different directions, therefore creating an active torque which is responsible for the rotation (see also discussion and conclusions, and Fig.~6). While we also monitor the average translational velocity, we find it is always zero (within numerical accuracy) in the whole range of activity explored in Fig.~1.

Zooming in close to the transition between the quiescent and the steadily rotating phases, we find evidence of an intermediate regime (between $\zeta=0.48\times 10^{-2}$ and $\zeta=0.53\times 10^{-2}$ included, Fig.~1(e) ) where the droplet oscillates, constantly switching between clockwise and anti-clockwise rotation. This behaviour leads to regular and large oscillations in the rotational velocity (Fig.~1(f), Suppl. Movie 2). While our simulations suggest that this switching may persist indefinitely, we cannot exclude the idea that a strong enough fluctuation may drive the system into the spontaneously rotating phase (as sometimes droplet switching precedes steady rotation for larger $\zeta$). When switching between the two senses of rotation, the droplet is maximally elongated and minimally bent. Our results show that a large enough activity can spontaneously break symmetry to select a given sense of rotation. Even so, as $\zeta$ increases we first observe a ``stop-and-go'' motion (e.g., for $\zeta=0.054\times 10^{-2}$, Suppl. Movie 3), and then a steadier motion where the amplitude of oscillations decreases with $\zeta$ (Fig.~1(g)).

Thus, extensile active droplets with strong normal anchoring can exist in one of three regimes: quiescent, oscillating, or steadily rotating, the intermediate oscillatory regime being stable for a small but finite range of activities. The transition between quiescent and oscillating states, and that between oscillations and steady rotations are both signalled by a singularity in the rotational velocity plot (Fig.~1(e)). Therefore, the onset of spontaneous rotation in active droplets appears to be qualitatively different from the onset of motility studied in Ref.~\cite{elsen}. For self-motile (translating) droplets, the transition is still associated with spontaneous symmetry breaking, yet it occurs without any intermediate regime in between the quiescent and spontaneously moving phases. Another important difference, of course, is that the self-motile droplets in Ref.~\cite{elsen} had no, or small, anchoring. 

\begin{figure}[htb]     
        \begin{center} 
        \includegraphics[width=0.5\textwidth]{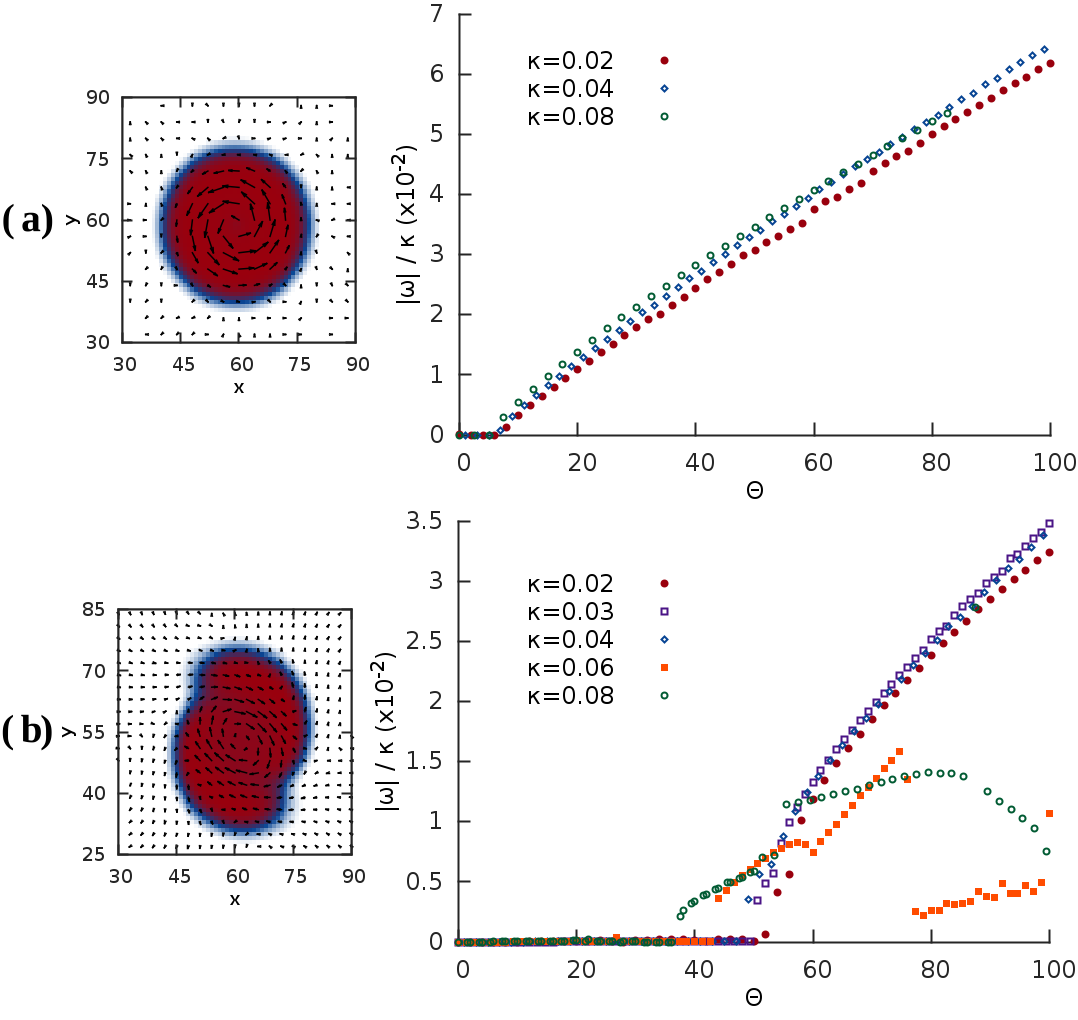}
\end{center}
\caption{Scaled plots of the rotational velocity over $\kappa$ versus $\Theta$. (a) Spiral initial condition: the curves approximately collapse onto each other, so that $\Theta$ is the single control parameter, as in Ref.~\cite{kruse}. (b) Aligned initial condition: there is now no scaling, apart from the case of smaller values of $\kappa$ and large $\zeta$ (i.e., deep in the spontaneously rotating phase). }
\label{fig2}                 
\end{figure}

It is relevant to ask here whether the mechanism leading to spontaneous rotation in Figure 1 is related to that discussed by Kruse {\it et al.} which leads to spontaneously rotating spiral defects~\cite{kruse,Jens}. In that case, rotations occur within a circular domain -- i.e., the surface tension is effectively infinite; also, again, there is no intermediate regime between the quiescent and the spontaneously rotating phases. Another important difference is that the theory in Ref.~\cite{kruse} does not include any anchoring: spontaneous rotation of these spirals is therefore triggered by activity alone, and hindered by elasticity -- correspondingly, the key dimensionless parameter was identified as $\Theta=\zeta R^2/\kappa$, which also regulates the transition between quiescent and self-motile droplets in Ref.~\cite{elsen}. Indeed, if we initialise our droplet as a spiral and slowly ramp up activity (Fig.~2(a)), we find that the threshold for rotation scales with elastic constants linearly, as expected.  Furthermore, all curves collapse when we plot $|\omega|/\kappa$ as a function of $\Theta$. However, this scaling is {\it not} obeyed for the case of aligned initial condition (Fig.~2(b)). Here, the threshold beyond which we observe spontaneous rotations does not simply scale as $\kappa$ (i.e., the transition points do not coincide in Fig.~2(b) when plotted versus $\Theta$). The deviation is especially visible for the larger values of $\kappa$ ($\kappa=0.06$ and $\kappa=0.08$). In these cases, the oscillating droplet regime is found for a larger range of activities; for $\kappa=0.08$, in particular, the stop-and-go rotation is also more prominent and leaves a signature in the rotational velocity curve so that there are three, rather than two, discontinuities. For the larger activities there are also less internal elastic deformations, and a much stronger active interfacial flow in the steadily rotating phase with respect to the cases of $\kappa=0.02, 0.04$.

Therefore, the mechanism leading to spontaneous rotation of our droplets is qualitatively different from that analysed in Ref.~\cite{kruse}, although they both require spontaneous symmetry breaking. The key difference is that here anchoring is important: for small anchoring, we see no spontaneous rotation. Furthermore, normal anchoring is required for extensile droplets to rotate. Indeed, we find no spontaneous rotations when we impose planar anchoring of the active dipole polarisation at the droplet surface; instead, we find that the droplet spontaneously moves as in Ref.~\cite{elsen}. It is important to note that the required combination of normal thermodynamic anchoring and extensile activity leads to a conflict between thermodynamic and active anchoring (as the latter is planar in this case~\cite{Matthew}).

Unlike the rotating spirals, our rotations are also typically associated with significant deformations of the droplet shape (see Fig.~1). Because our mechanism for spontaneous rotations requires a certain type of anchoring, with a sufficiently large strength, we refer to them in what follows as ``anchoring-driven rotations''.

\subsection{Spontaneous rotation of contractile droplets}

We next examine the case of a contractile droplet -- recall this is a simple model for an actomyosin droplet. Contractile droplets can also display spontaneous anchoring-driven rotations; interestingly, this time the required anchoring is planar. Once again, we highlight that this anchoring is required to contrast the active-induced anchoring, which for contractile gels is normal~\cite{Matthew}.

Figure 3 lists the four main possible regimes which we find for contractile droplets in the parameter range which we have explored. First, if the initial condition is an aster (or spiral), flow-tumbling droplets (but not flow-aligning ones) are spontaneously rotating (Fig.~3(a)). This is the analogue of the rotating aligning spirals for extensile droplets. For an aligned initial condition, instead, we find three possible steady state scenarios. For small activity, there is a quiescent phase, where the droplet is stationary, even if there is active flow inside (Fig.~3(b)). This is the analogue of the quiescent phase in the extensile case (note that the droplet now extends perpendicular, rather than parallel, to the direction of polarisation). Second, for large enough activity, the droplet deforms into an S-shape (Fig.~3(c)). In this regime, the droplet spontaneously rotates. The shape of the deformed droplet resembles that observed in the extensile case -- this is therefore the contractile analogue of the anchoring-driven rotating regime (Fig.~3(c), Suppl. Movie 4). In this regime, the internal flow has a vortex-like shape, again in analogy with the extensile case -- now, the active torque is provided by splay, rather than bend, deformation patterns. Finally, for even larger activity (Fig.~3(d), Suppl. Movie 5), and still for finite anchoring, we observe an asymmetric droplet which is revolving around a point distinct from its centre of mass.

\begin{figure}[htb]     
        \begin{center} 
        \includegraphics[width=0.45\textwidth]{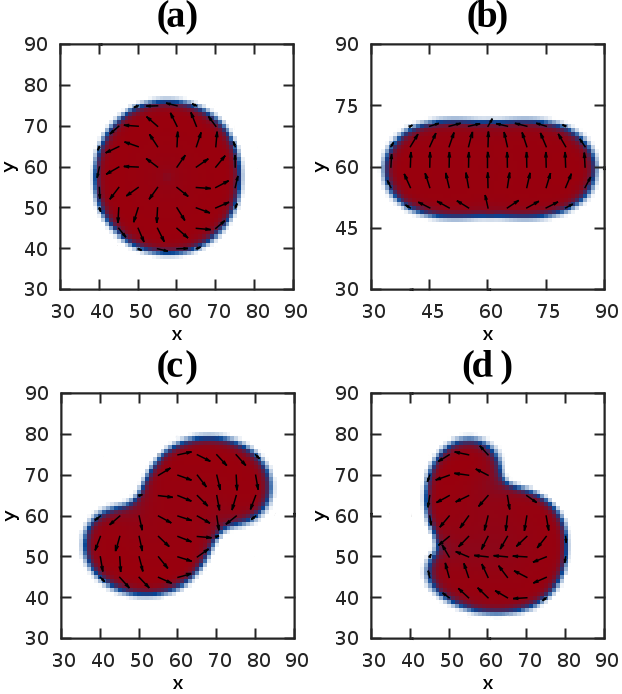}
\end{center}
\caption{Snapshots of droplet shapes and internal polarisation in the four main regimes found for contractile droplets with planar anchoring. (a) Rotating spiral. (b) Quiescent droplet: note the deformation perpendicular to the polarisation field. (c) Steadily rotating symmetric droplet. (d) Steadily rotating asymmetric droplet: the centre of rotation is distinct from the centre of mass. The polar nematic in (a) is flow-tumbling; in (b)-(d) is flow aligning.}
\label{fig3}                 
\end{figure}

To characterise the transition between quiescent and spontaneously rotating phases for contractile droplets more quantitatively, in Fig.~4 we show the (modulus of the) rotational velocity as a function of activity: this should be compared with the analogous plots in Figures 1,2 for extensile droplets. First, we plot in Figure 4(a) the results obtained for a spiral initial condition. In order to find rotations with this initial condition, we need to consider a flow-tumbling, rather than flow-aligning material. As for extensile rotating spirals, there is a single discontinuity in the rotational velocity curve (Fig.~4(a)).

Next, we consider the case of aligned initial conditions (Fig.~4(b)). To generate the rotational velocity curve shown in Figure 4(b), we simulated a hysteresis loop whereby the activity is first ramped up and then turned down: in practice, each simulation is initialised with the steady state conformation obtained in the run with the neighbouring value of the activity. The fact that there is hysteresis suggests that the transition between the quiescent and rotating phase is a discontinuous, first-order-like, one. With respect to the extensile case, there is no oscillatory droplet regime, but instead a window of activities where the droplet either steadily rotates (in the top branch in Fig.~4(b)) or does so only transiently and is static at later times (in the bottom branch in Fig.~4(b)). The additional singularity at larger $\zeta$ corresponds to the transition between the symmetric and asymmetric rotating droplets (Figs.~3(c) and 3(d) respectively).

\begin{figure}[!h]     
        \begin{center} 
        \includegraphics[width=0.45\textwidth]{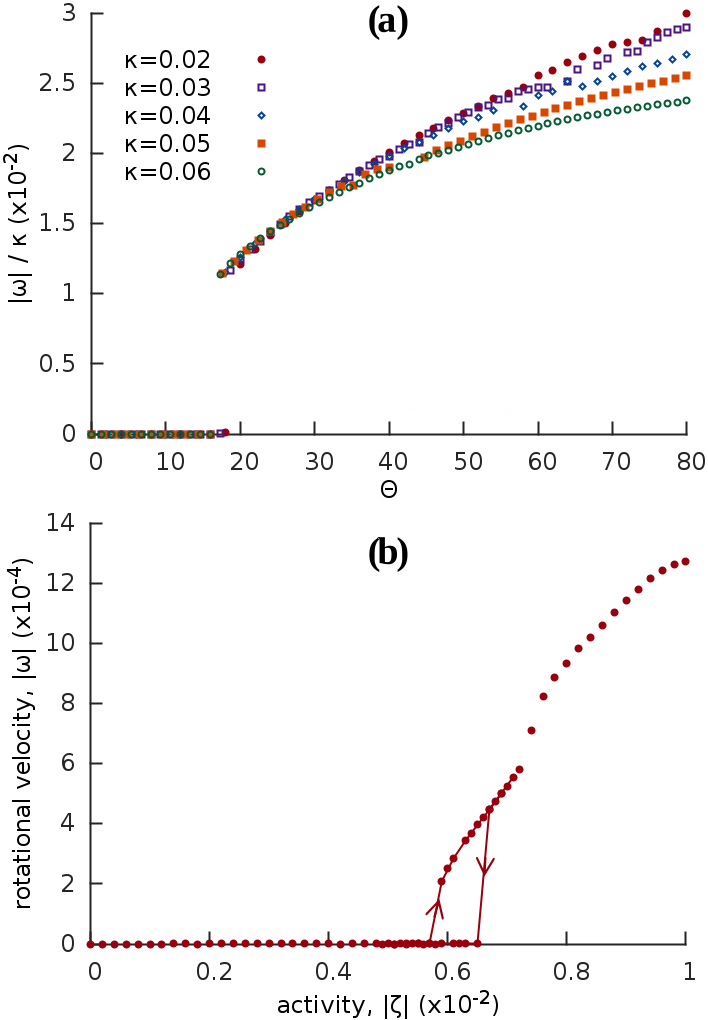}
\end{center}
\caption{Plot of the rotational velocity as a function of activity for: (a) a contractile flow-tumbling droplet with a spiral initial condition (the rotational velocity is rescaled by $\kappa$, and the control parameter is the dimensionless activity $\Theta$); (b) a contractile flow-aligning droplet with an aligned initial condition.}
\label{fig4_v0}                 
\end{figure}

Previous work in quasi-1D active nematics~\cite{Julia_UnifiedPicture} showed that extensile and contractile systems can be exactly mapped into each other by a simultaneous change in sign of $\zeta$ and $\xi$, accompanied by a rotation of the polarisation field by $\pi/2$. In this work contractile and extensile droplets behave differently as we study rodlike systems ($\xi>0$) in both cases. Still, there is an approximate mapping between extensile droplets with normal anchoring and contractile ones with planar anchoring, and between extensile droplets with planar anchoring and contractile ones with normal anchoring (as discussed in Fig.~6 below).

\subsection{The significance of the anchoring}

As highlighted above, our spontaneous rotations are anchoring-driven: extensile rotations require normal anchoring, contractile ones require planar anchoring. Besides the nature of the anchoring, it is important to characterise the dependence on its strength, and that is what we do in this Section. 

\begin{figure}[!h]     
        \begin{center} 
        \includegraphics[width=0.45\textwidth]{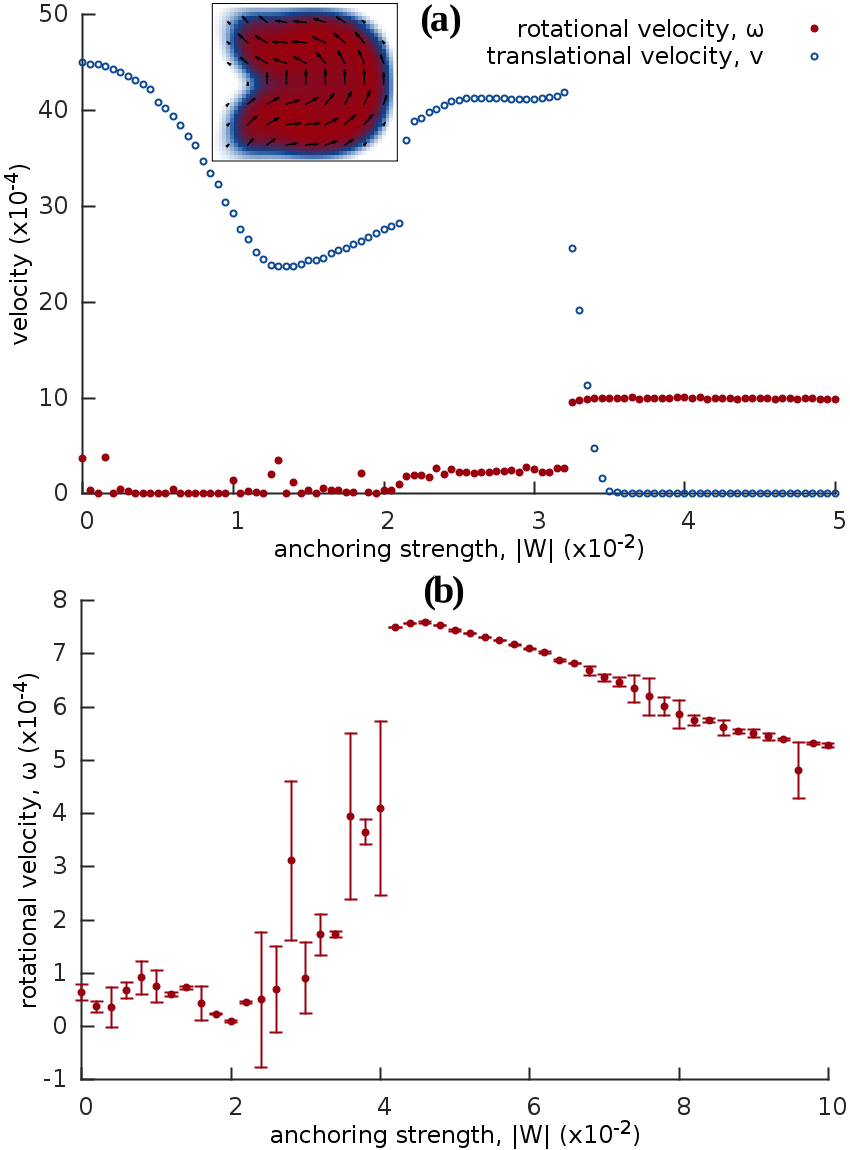}
\end{center}
\caption{(a) Plot of the rotational and translational velocities for an extensile droplet with variable strength, $W$, of normal anchoring. For small $W$, the droplet mainly translates, whereas it mainly rotates for large $W$. (b) Plot of the rotational velocity for a contractile droplet with planar anchoring. In this case fluctuations are larger, hence we have averaged over 10 realisations (each initialised with a different random polarisation pattern). Error bars represent the standard deviation associated with each average calculation. }
\label{fig5}                 
\end{figure}

We first consider the case of extensile droplets. Figure 5(a) plots the rotational and translational velocities as a function of the strength of normal anchoring, $W$, for a fixed value of the extensile activity (large enough to create spontaneous rotations in Fig.~1). If $W$ is small, we observe little or no rotation: in this regime there is still motion, but in the form of translations. This is the regime of self-motile droplets studied in Ref.~\cite{elsen}, where bend deformations drive active flows which power the motion of the droplet. Activity drives an effective planar anchoring at the surface (see left snapshot in Fig.~5(a)). If $W$ increases past a critical threshold, translatory motion arrests and we enter the previously studied spontaneously rotating regime (Fig.~1(c)). 

These results therefore suggest that in order to observe anchoring-driven rotations we need a large enough value of both $\Theta$ and $W$. This is in contrast to both the self-motile translating droplets of Ref.~\cite{elsen} and the rotating spirals of Ref.~\cite{kruse} which solely require a large value of $\Theta$.

Next, we analyse the case of contractile droplets with planar anchoring (Fig.~5(b)). These simulations confirm the qualitative trend found in the extensile, as rotations only occur for a sufficiently large value of $W$. However, here the picture is more complex. First, in the rotating regime, as previously discussed, the droplet may develop a shape asymmetry (Fig.~3(d)), which leads to a non-zero value of the translational velocity as well. Second, in the non-rotating phase, with the parameters in Fig.~5(b) we observe irregular motion, rather than steady translation as for the extensile case (Fig.~5(a)). This is possibly due to the smaller value of the surface tension used in Fig.~5(b) (with respect to Fig.~5(a)). As a result of this choice, which is due to reasons of numerical stability, the droplet largely deforms in the low $W$ regime, and can accomodate multiple splay elastic deformations which are likely to be responsible for the more complex dynamics.  The contractile results also highlight a large variability in the observed rotational and translational velocity, due to different random noise in the initial condition.

A possible explanation of these results is that, as previously hinted, droplet rotations require a thermodynamic anchoring which conflicts with the active anchoring (see cartoon in Fig.~6). Consider for concreteness the case of an extensile droplet. In the absence of any conflict, or when active anchoring wins, the droplet can settle into a structure consisting of a single bend deformation, whose geometry is similar to that of the droplets in Ref.~\cite{elsen} and consistent with steady translation (but no rotation, see Fig.~6(a)). Instead, in the presence of a conflict, the anchoring can stabilise a different pattern of polarisation inside the droplet, with two bend deformations of opposing sense: this in turn can create the torque required for the active rotation (Fig.~6(b)). Flipping the sign of $\zeta$, and replacing bend with splay, a similar reasoning explains the behaviour of contractile droplets (Figs.~6(c),(d)). 

Consequently, we speculate that an important dimensionless number to describe rotations should be the ratio between the effective thermodynamic and active anchoring. The effective thermodynamic anchoring can be estimated as $w_t=W/l_i$, where $W$ is the anchoring strength entering in our free energy density (which has units of N), and $l_i$ is the interfacial thickness (see Model and methods). The active anchoring can instead be estimated as $w_a\sim \zeta l_a$, where $l_a$ is an active lengthscale characterising the size of splay-bend distortion within an active nematic~\cite{giomi}. A more quantitative test of this expectation is here hampered by the fact that the droplet behaviour does not only depend on $w_t/_a$, but also on $\Theta$ and the active capillary number.

\section{Discussion and conclusions}

In summary, we have shown here that active droplets with sufficiently strong thermodynamic anchoring show a transition between a quiescent phase and one where droplets rotate spontaneously. This behaviour contrasts with that of active droplets without thermodynamic anchoring, where the transition is instead between a quiescent and a self-motile (and non-rotating) phase~\cite{elsen,giomi}. Our spontaneously rotating droplets are fundamentally distinct from the rotating spirals of~\cite{kruse,kruselong}, as the latter arise even with no anchoring at all, and infinite surface tension.

\begin{figure}
\begin{center}
\begin{tabular}{ c | c }
\hline
Initial condition \\and anchoring & Steady state
\\
\hline
\\
 \bf{(a)}
 \includegraphics[align=c, width=0.20\textwidth]{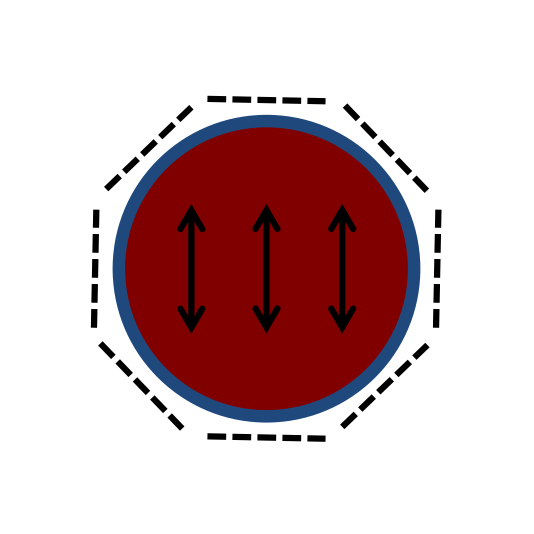} & 
 \includegraphics[align=c, width=0.20\textwidth]{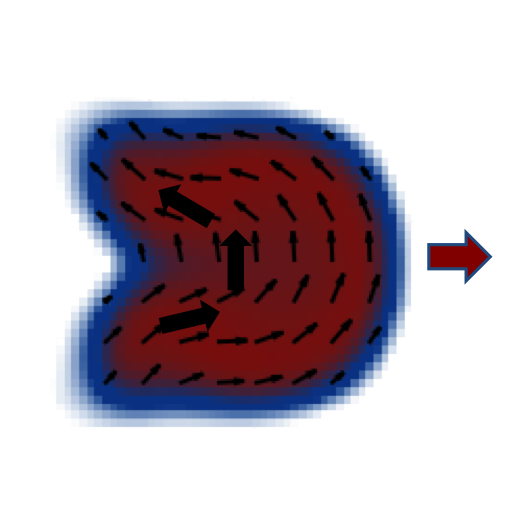} \\

\hline
\\
\bf{(b)}
 \includegraphics[align=c, width=0.20\textwidth]{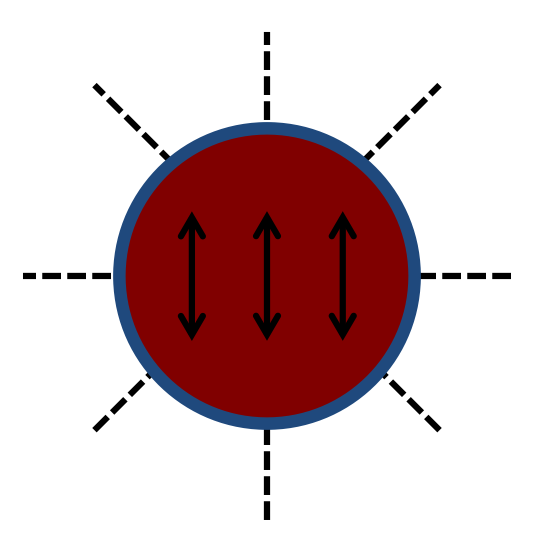} & 
 \includegraphics[align=c, width=0.20\textwidth]{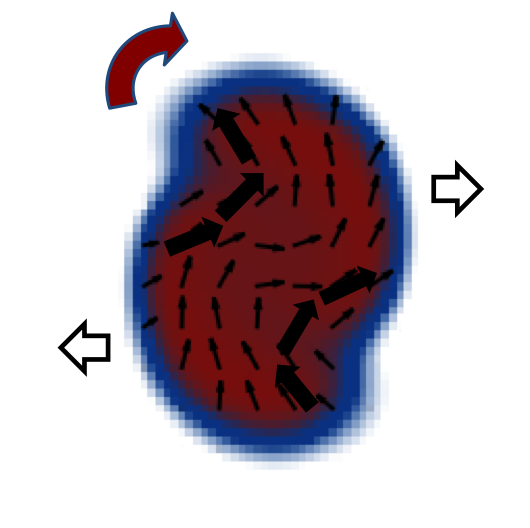} \\

\hline
\\
\bf{(c)} 
\includegraphics[align=c, width=0.20\textwidth]{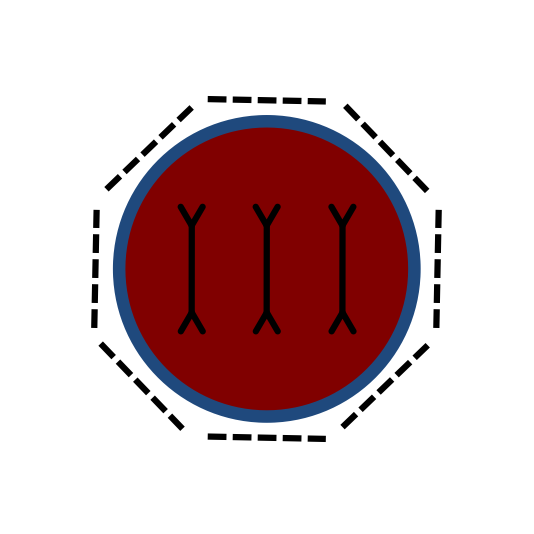} & 
\includegraphics[align=c, width=0.20\textwidth]{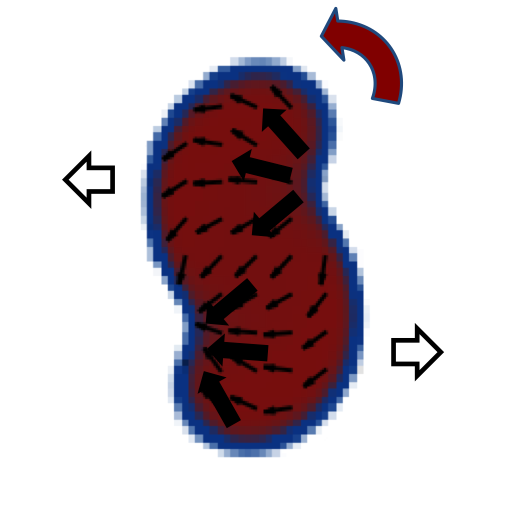} \\

\hline
\\
\bf{(d)} 
\includegraphics[align=c, width=0.20\textwidth]{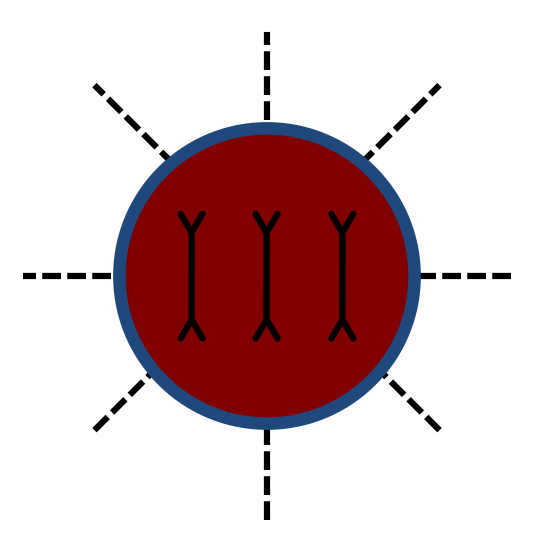} & 
\includegraphics[align=c, width=0.20\textwidth]{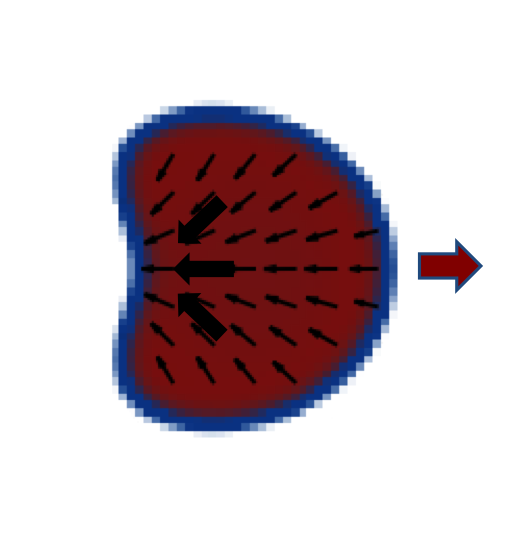} \\

\hline 

\end{tabular}

\end{center}
\caption{This cartoon illustrates the interplay between anchoring and activity to determine the dynamics of our active gel droplets. Extensile droplets (a) and (b)) rotate with normal anchoring (b), because this anchoring stabilises a pair of bend deformations which creates a sustained active torque (here leading to clockwise rotation). Instead, planar anchoring or no anchoring (represented as dashed lines in (a)) leads to a single bend deformation, a pattern compatible with translational motility. Similar considerations apply to contractile activity (rows (c) and (d), see text). Red arrows represent the overall sense of movement while black and white arrows represent flows generated by each deformation. } 
\label{cartoon}  
\end{figure}

A schematic representation of the mechanism leading to the anchoring-driven rotations we observe is drawn in Figure~\ref{cartoon}. For instance, in extensile droplets normal anchoring stabilises two regions of bend distortions: these are associated with two local active forces with opposing direction. When the two forces are not collinear, for instance due to a fluctuation, they create an active torque which rotates the droplet. The elastic distortions are accompanied by deformations in the surface. Importantly, the anchoring is required to maintain two symmetric regions of bend: when it is absent, there is only one deformation in steady state, and the droplet translates (Fig.~6(b)).  Similar reasoning explains qualitatively why contractile droplets with planar anchoring also rotate (Fig.~\ref{cartoon}).

The morphology and kinetics of our spontaneously rotating droplets is very varied, and can be controlled by a number of parameters. For instance, we can slow down or arrest rotation by decreasing anchoring strength. Close to the transition between rotating and motile droplets, we also find more exotic dynamics: an example is given in Suppl. Movie 6 where the droplet alternates between motile spells and partial rotations. Surface tension and elasticity have a further important effect, as they control the pattern of the active flow, which can be maximal in the droplet interior or at the surface. 

An important question is what a suitable system to recreate these results in the lab might be. We suggest that a good candidate may be an emulsion incorporating the hierarchically assembled active matter described in~\cite{sanchez}. This system is made up by oil-water emulsions where the water component contains suspensions of microtubule, polyethylene glycol (PEG) and kinesin motors. Due to the presence of PEG, there are depletion interactions which cause microtubules to stick to each other to form bundles, and motor activity causes such bundles to slide antiparallel to each other, creating extensile active dipolar forces. A difference with our model is that in the experimental system the order is nematic, rather than polar. However, by carrying out simulations of an active droplet where the active phase is nematic (the simulation method is the same which was discussed in Ref.~\cite{Matthew}), we have found that such spontaneous rotations occur in that case as well, with only minor differences to the case reported here (see Suppl. Movies 7 and 8). Therefore, the physics we observe here should in principle apply to the apolar material described in Ref.~\cite{sanchez}. Still, a number of practical challenges would need to be addressed before spontaneously rotating droplets can be self-assembled, as experimental studies in the literature report the formation of an active microtubule-kinesin shell, rather than a uniformly filled droplet, and the natural anchoring in those extensile systems normally is planar -- this is favoured as a consequence of the microtubule rigidity. Notwithstanding these issues, we hope that our results will stimulate experiments on the self-assembly of active soft rotators in the future, based on either active gels or alternative designs. 

We thank Michael E. Cates for useful discussions and acknowledge EPSRC (grant EP/J007404/1) for support.

\end{document}